**TYMOSHCHUK Dmytro**
Ternopil Ivan Puluj National Technical University
https://orcid.org/0000-0003-0246-2236
e-mail: dmytro.tymoshchuk@gmail.com

**YATSKIV Vasyl**
West Ukrainian National University
https://orcid.org/0000-0001-9778-6625
e-mail: jazkiv@ukr.net

# USING HYPERVISORS TO CREATE A CYBER POLYGON

*Cyber polygon used to train cybersecurity professionals, test new security technologies and simulate attacks play an important role in ensuring cybersecurity. The creation of such training grounds is based on the use of hypervisors, which allow efficient management of virtual machines, isolating operating systems and resources of a physical computer from virtual machines, ensuring a high level of security and stability. The paper analyses various aspects of using hypervisors in cyber polygons, including types of hypervisors, their main functions, and the specifics of their use in modelling cyber threats. The article shows the ability of hypervisors to increase the efficiency of hardware resources, create complex virtual environments for detailed modelling of network structures and simulation of real situations in cyberspace.*
*Keywords: Hypervisor, cyber polygon, VMware ESXi, Microsoft Hyper-V, Xen, KVM, Cluster.*

**ТИМОЩУК Дмитро**
Тернопільський національний технічний університет імені Івана Пулюя

**ЯЦКІВ Василь**
Західноукраїнський національний університет

# ВИКОРИСТАННЯ ГІПЕРВІЗОРІВ ДЛЯ СТВОРЕННЯ КІБЕРПОЛІГОНУ

*Кіберполігони, які використовуються для тренування фахівців з кібербезпеки, тестування нових технологій захисту та моделювання атак, відіграють важливу роль у забезпечення кібербезпеки. В основі створення таких полігонів лежить використання гіпервізорів, які дозволяють ефективно управляти віртуальними машинами, ізолюючи операційні системи та ресурси фізичного комп'ютера від віртуальних машин, забезпечуючи високий рівень безпеки та стабільності. В роботі проаналізовано різні аспекти використання гіпервізорів у кіберполігонах, включаючи типи гіпервізорів, їхні основні функції, а також специфіку застосування у моделюванні кіберзагроз та розгортанні кіберполігонів. Показано здатність гіпервізорів підвищувати ефективність використання апаратних ресурсів, створювати складні віртуальні середовища для детального моделювання мережевих структур та імітації реальних ситуацій в кіберпросторі.*
*Ключові слова: Гіпервізор, кіберполігон, VMware ESXi, Microsoft Hyper-V, Xen, KVM, Cluster.*

## INTRODUCTION

Information technology is constantly evolving, creating new challenges for cybersecurity. Cyber polygon, or environments for modelling cyber threats, are becoming a necessary tool for training specialists, testing system security, and researching new methods of countering attacks. The basis for creating such training grounds is the use of hypervisors.

A hypervisor is software that manages the operation of virtual machines. It allows you to create and effectively manage virtual machines, each with its own operating system and applications.

A cyber polygon is a virtual environment used to train cybersecurity professionals, test new security technologies, simulate attacks, and practice incident response scenarios. It can include virtual machines, network equipment, storage systems and other components that simulate real-world infrastructure.

## MAIN PART

The hypervisor distributes physical resources (such as CPU, memory, disk space, etc.) among several virtual machines, ensuring their independence and stability. This makes it possible to run several different operating systems on the same physical server, which increases the flexibility and efficiency of using hardware resources. The hypervisor ensures that virtual machines cannot interact with each other in an undesirable way or affect the host system, which provides a high level of security and stability. It can include various functions, such as managing virtual networks, data storage, and I/O processing. Thanks to its architecture, a hypervisor allows you to create and manage virtualised environments that can be used for a variety of purposes.

There are two types of hypervisors: type 1 (bare-metal) and type 2 (hosted).

Type 2 runs on top of an operating system that is already installed on a physical server (Figure 1). Examples: VMware Workstation, Oracle VirtualBox, Parallels Desktop, VMware Fusion, UTM, QEMU

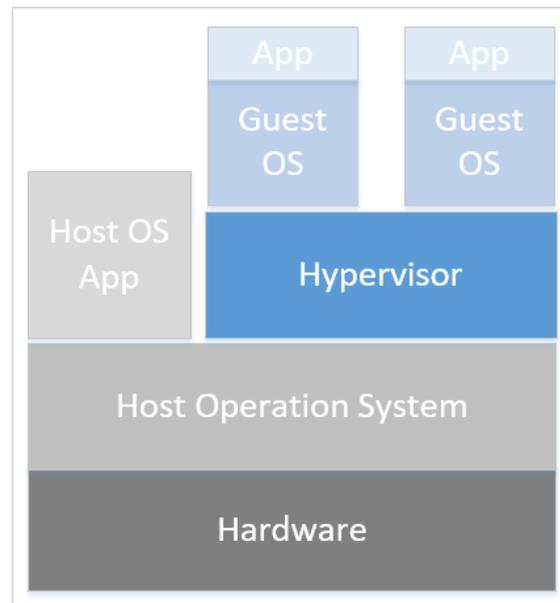

**Fig. 1 Type 2 hypervisor architecture**

Type 1 runs directly on the hardware without the need for an underlying operating system (Figure 2). Examples: VMware ESXi, Microsoft Hyper-V, Xen, and KVM.

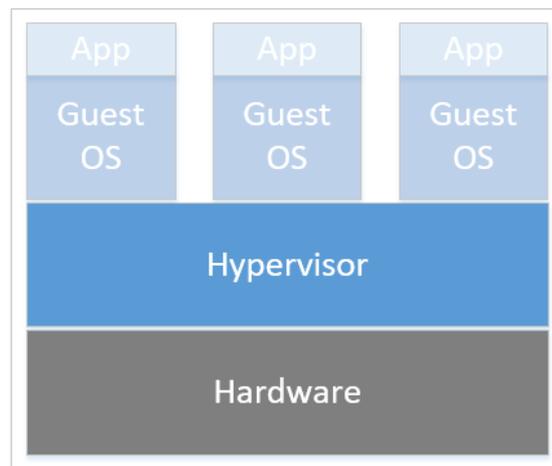

**Fig. 2 Type 1 hypervisor architecture**

VMware ESXi is a high-performance Type 1 hypervisor that is small and fast. The hypervisor delivers high performance while minimising virtualisation overheads thanks to its lightweight architecture [1]. High level of scalability. ESXi allows you to efficiently use server resources by running many virtual machines on a single server and automatically managing them through vCenter Server, allowing you to create large virtualised environments.

One of the main advantages of VMware ESXi is its high reliability. The hypervisor supports high availability features such as High Availability and Fault Tolerance. The vMotion function allows you to move virtual machines from one server to another without interrupting their operation, which is important for ensuring the smooth operation of services. The DRS (Distributed Resource Scheduler) tool automatically balances the load between servers by moving virtual machines according to available resources. Storage vMotion technology allows you to move virtual machines between different datastores without interrupting their operation. VMware vSAN software-defined storage system allows you to create virtualised storage based on local server disks. vSAN combines disc resources from multiple servers into a single cluster, creating scalable and high-performance storage for virtual machines.  All of these are part of the VMware vSphere platform functionality and are critical technologies for ensuring the continuous operation of virtualised environments. VMware ESXi supports both Intel VT (Intel Virtualisation Technology) and AMD-V (AMD Virtualisation).

Intel VT and AMD-V are hardware-based virtualisation technologies developed by Intel and AMD, respectively. Intel VT and AMD-V are critical components for virtualisation performance in modern hypervisors. They leverage these hardware capabilities to improve virtualisation performance, security, and efficiency.

VMware ESXi can provide a solid foundation for a cyber polygon with its high-performance virtualisation capabilities, scalability, and reliability. It allows you to create complex and resource-intensive virtual environments that can mimic real-world network structures with different types of servers, operating systems, and applications. This allows you to configure various cyber threat and attack scenarios that can be tested without risking real systems. Using features such as vSphere vMotion and VMware vSAN, you can ensure that the cyber range is always running, even when performing updates or moving virtual machines between servers, which is important for long training and education sessions. VMware ESXi allows you to create complex network topologies with different network segments and access policies. This allows you to recreate real-world enterprise network conditions, including data distribution and protection. VMware ESXi also supports snapshots of virtual machines, which allows you to quickly return to previous system states after tests or attacks.

Hyper-V is a Type 1 hypervisor developed by Microsoft that allows you to create and manage virtual machines [2]. It is part of the Windows Server and Windows 10/11 (Pro, Enterprise, and Education) operating systems, providing users with the ability to efficiently virtualise servers and workstations. Hyper-V directly controls resources such as CPU, memory, and network interfaces, distributing them among virtual machines. Live Migration allows you to move running virtual machines between physical servers without interrupting their operation. Hyper-V Replica provides replication of virtual machines between two different sites, allowing for backup and disaster recovery. Despite its many benefits, Hyper-V does have some limitations. Although Hyper-V supports a variety of operating systems, some versions of Linux or other operating systems may have limited support or require additional configuration.

The Hyper-V hypervisor can serve as a solid foundation for a cyber polygon with its powerful virtualisation capabilities, integration with other Microsoft products, and flexibility in managing virtual environments. With Hyper-V, you can easily create isolated virtual environments in which to deploy a variety of operating systems and applications to simulate and defend against attacks. Integration with Active Directory allows you to control access to resources and manage users. Hyper-V also supports replication and backup functions, which allows you to preserve the state of your environment, quickly restore it after test attacks, or create backups for analysis. Hyper-V virtual switches allow you to configure complex network topologies by simulating various types of network interactions. With Hyper-V, you can implement a scalable cyber polygon that can be used for both individual training sessions and complex team exercises.

Xen is an open-source Type 1 hypervisor developed at the University of Cambridge and is one of the most popular virtualisation solutions in the world [3]. It allows you to run multiple operating systems on a single physical server, efficiently sharing hardware resources between them. Xen runs directly on the server hardware, providing a high level of performance and security by isolating virtual machines from each other.

One of the key features of Xen is its ability to support different virtualisation models, including paravirtualisation and hardware virtualisation. Paravirtualisation allows operating systems running on Xen to be optimised for virtualisation, which improves performance. Hardware virtualisation, on the other hand, allows you to run unmodified operating systems such as Windows through the use of special hardware extensions of the Intel VT or AMD-V processor. Xen supports high availability and live migration of virtual machines, allowing them to be moved between physical servers without interruption. Xen provides a command-line interface and tools for managing virtual machines, allowing administrators to effectively monitor and configure the virtualised environment.

Xen can be a good foundation for creating a cyber polygon due to its flexibility, open source, and ability to support different virtualisation models. Since Xen supports both paravirtualisation and hardware virtualisation, it allows you to create cyber polygons with different operating systems, including virtualisation-optimised versions of Linux and unmodified versions of Windows. Xen's flexibility allows you to create complex network topologies and scenarios to simulate real-world attacks. With support for live migration of virtual machines and high availability, Xen ensures the continuity of the cyber range, even during updates or the transfer of virtual machines to other physical servers. This allows you to conduct long-term training and testing without the risk of stopping the environment.

KVM (Kernel-based Virtual Machine) is a Type 1 hypervisor that is part of the Linux kernel and allows you to use hardware virtualisation features to create and manage virtual machines [4]. As a virtualisation solution, KVM delivers high performance and efficiency through tight integration with the Linux kernel, allowing it to use existing system components and resources. The main component of KVM is the kernel module, which turns the Linux kernel into a hypervisor that manages virtual machines.

One of the key advantages of KVM is that it uses standard Linux tools and mechanisms to manage virtual machines. This includes tools for monitoring, network configuration, and resource management. KVM supports a variety of operating systems, including different versions of Linux, Windows, and other OSes, making it a versatile virtualisation solution.

KVM integrates with other projects and technologies, such as libvirt, which provides interfaces for managing virtual machines via the command line or graphical interfaces.

Another important part of KVM is QEMU. QEMU (Quick Emulator) is a versatile open source emulator that allows you to create virtual machines by emulating various hardware [5].

QEMU provides emulation of a wide range of hardware components, such as processors, network cards, hard drives and other peripherals. One of the main advantages of QEMU is its ability to support different processor architectures such as x86, ARM and others. QEMU can run in both full emulation mode and hardware virtualisation mode to improve performance. In full emulation mode, QEMU provides emulation of all hardware, allowing you to run virtual machines regardless of whether the physical host supports hardware virtualisation.

KVM is an excellent basis for creating a cyber polygon due to its integration with the Linux kernel and powerful virtualisation capabilities. As a specialised environment for modelling and analysing cyber threats, a cyber polygon requires a stable, scalable and reliable platform for deploying virtual machines and simulating attacks. KVM, as part of the Linux kernel, provides high performance and efficiency through the use of hardware virtualisation via Intel VT or AMD-V. This allows you to create virtual machines with minimal overhead, which is especially important for large and resource-intensive environments such as cyber polygons. QEMU, when used in conjunction with KVM, provides the ability to emulate different hardware, allowing you to create virtual machines with different characteristics and configurations. This is useful for modelling various attack and defence scenarios within a cyber polygon. QEMU also supports virtual machine snapshots, which allows you to save system states and quickly restore them if necessary. The migration capability also helps to ensure high availability and scalability of the cyber polygon, adapting it to changing needs.

Type 1 hypervisors, which work directly with the hardware, provide higher performance than type 2 hypervisors because they do not require an underlying operating system. This is critical for cyber polygons, where high data processing speeds and realistic modelling of cyber threats are required. Due to these advantages, Type 1 hypervisors are ideal for creating a cyber polygon, providing a high level of performance, security, scalability and stability that are critical for effective testing and modelling of cyber threats.

One of the most important aspects of creating a cyber polygon is the use of routers with firewall functionality in the test environment. Routers such as the Cisco CSR1000v, IPFire, OPNsense, pfSense, MikroTik CHR, Juniper Cloud-Native Router, FortiGate VM, and Palo Alto VM-Series are a solid choice for a cyber polygon due to their advanced security, configuration, and integration with other systems [7][8][9][10][11][12][13]. Using these routers with firewall capabilities allows you to create a variety of scenarios for testing and training in the cyber lab. Their capabilities help to create complex and realistic network environments that allow you to effectively model and analyse cyber threats.

Figure 3 shows an architectural diagram of a cyber polygon created on the basis of a cluster of hypervisors.

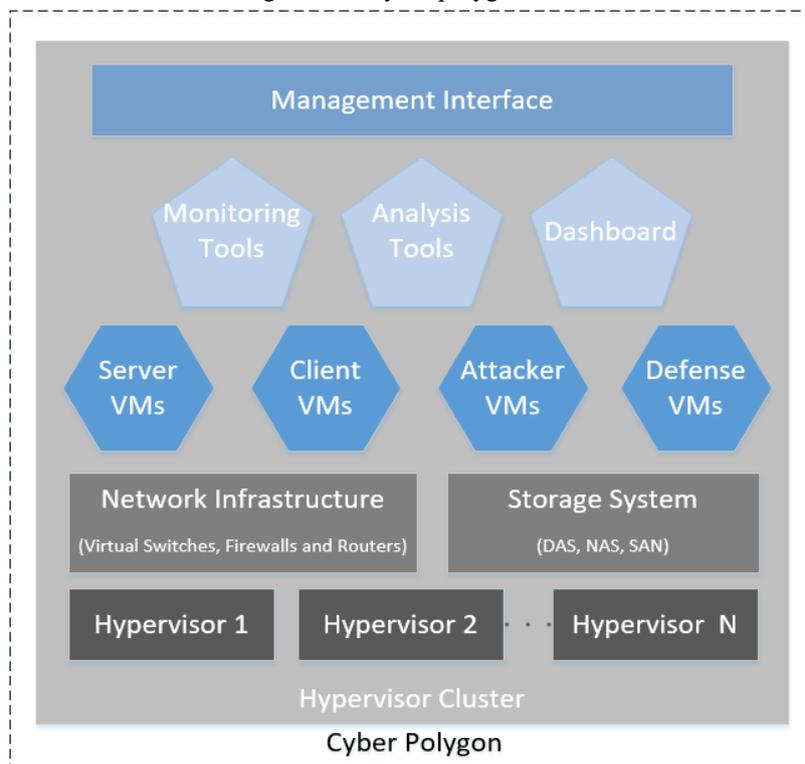

**Fig. 3 The architecture of the cyber polygon**

The centralised management interface allows you to manage all virtual machines, configure network configurations, and monitor the status of the entire hypervisor cluster. A hypervisor cluster consists of several physical servers with hypervisors. There are several types of virtual machines in the architecture: server machines that simulate various server roles, such as web servers, databases, file servers; client machines that simulate user endpoints, such as computers and mobile devices; attack machines configured with tools and scripts to perform cyber attacks; and machines that are equipped with security tools to detect and mitigate attacks. The network infrastructure includes virtual switches, routers and firewalls that control and monitor data flows between virtual machines and separate network segments. The storage system includes DAS (Direct-Attached Storage), NAS (Network-Attached Storage), and SAN (Storage Area Network), which provide data storage for virtual machines. Monitoring, analysis tools and a dashboard allow real-time monitoring and analysis of network activity and incident response by visualising the security status.

This architecture allows for the creation of diverse and realistic scenarios for cybersecurity training, providing the flexibility and scalability to conduct complex tests in a controlled environment.

## CONCLUSIONS

Using a Type 1 hypervisor to create a cyber polygon is an effective approach that allows you to simulate complex and realistic cyber threat scenarios. The main advantages are high performance, reliability and isolation of virtual machines, which creates a secure environment for testing and training. Type 1 hypervisors provide the ability to create scalable and adaptive cyber polygons that can be used to train cybersecurity professionals, conduct research and test new security technologies.

Thus, a Type 1 hypervisor is a key element for creating effective and secure cyber polygons that can be used in various industries, from educational institutions to corporate environments.

## References

1. VMware vSphere Documentation. URL: https://docs.vmware.com/en/VMware-vSphere/index.html
2. Hyper-V technology overview. URL: https://learn.microsoft.com/en-us/windows-server/virtualization/hyper-v/hyper-v-technology-overview
3. Xen project software overview. URL: https://wiki.xenproject.org/wiki/Xen_Project_Software_Overview#PV_.28x86.29
4. Kernel Virtual Machine. URL: https://linux-kvm.org/page/Main_Page
5. QEMU documentation. URL: https://www.qemu.org/docs/master/
6. Introduction – OPNsense documentation. URL: https://docs.opnsense.org/intro.html
7. IPFire documentation. URL: https://www.ipfire.org/docs
8. Introduction pfSense Documentation. Netgate Documentation. URL: https://docs.netgate.com/pfsense/en/latest/general/index.html
9. Cisco Cloud Services Router 1000v. URL: https://www.cisco.com/c/en/us/support/routers/cloud-services-router-1000v/model.html
10. Cloud hosted router - mikrotik documentation. URL: https://help.mikrotik.com/docs/display/ROS/Cloud+Hosted+Router,+CHR
11. Juniper Cloud-Native Router (JCNR). URL: https://www.juniper.net/documentation/product/us/en/juniper-cloud-native-router/
12. FortiGate private cloud 7.4. Fortinet Document Library. URL: https://docs.fortinet.com/product/fortigate-private-cloud/7.4
13. VM-Series. Palo Alto Networks Tech Docs Home. URL: https://docs.paloaltonetworks.com/vm-series